\documentclass[prl,twocolumn,showpacs,preprintnumbers,amsmath,amssymb,superscriptaddress]{revtex4}
\usepackage{graphicx}
\usepackage{dcolumn}
\usepackage{bm}

\begin{document}

\title{Metamagnetic Transition in Na$_{0.85}$CoO$_2$ Single Crystal}

\author{J. L. Luo, N. L. Wang, G. T. Liu, D. Wu, X. N. Jing}

\affiliation{Beijing National Laboratory for Condensed Matter
Physics, Institute of Physics, Chinese Academy of Sciences,
Beijing 100080, China}

\author{F. Hu, T. Xiang}

\affiliation{Institute of Theoretical Physics and
Interdisciplinary Center of Theoretical Studies, Chinese Academy
of Sciences, P. O. Box 2735, Beijing 100080, China}

\date{\today}

\begin{abstract}
We report the magnetization, specific heat and transport
measurements of a high quality Na$_{0.85}$CoO$_2$ single crystal
in applied magnetic fields up to 14T. In high temperatures, the
system is in a paramagnetic phase. It undergoes a magnetic phase
transition below $\sim 20$K. For field $H\parallel c$, the
measurement data of magnetization, specific heat and
magnetoresistance reveal a metamagnetic transition from an
antiferromagnetic state to a quasi-ferromagnetic state at about 8T
in low temperatures. However, no transition is observed in the
magnetization measurements up to 14T for $H\perp c$. The low
temperature magnetic phase diagram of Na$_{0.85}$CoO$_2$ is
determined.
\end{abstract}


\maketitle

The recent discovery of superconductivity in the hydrated cobalt
oxides \cite{Terasaki, Takada, Yayu} has attracted extensive
attention, since it is the only known layered transition metal
oxide, which exhibits superconductivity other than high $T_c$
cuprates and Sr$_2$RuO$_4$ \cite{Maeno}. The Na$_x$CoO$_2$ crystal
consists of two-dimensional CoO$_2$ layers separated by Na layers,
similar to that of high $T_c$ cuprates and Sr$_2$RuO$_4$ except
that in each layer Co atoms form a triangular lattice rather than
a square one \cite{Lynn, Jorgensen}. Superconductivity is observed
when x $\sim$ 0.3 and sufficient water is intercalated between the
CoO$_2$ layers. A number of experiments \cite{Waki, Fujimoto,
Maska} and theoretical works \cite{Hu, Baskaran} suggested that
superconductivity in Na$_{x}$CoO$_2$.H$_2$O is unconventional,
probably with spin triplet pairing.

With change of Na content x, Na$_{x}$CoO$_2$ exhibits a rich phase
diagram. Foo et al. studied transport and magnetic properties of a
series of Na$_{x}$CoO$_2$ samples with x varying from 0.3 to 0.75
\cite{Foo}. They found that the ground state is a paramagnetic
metal at x$\sim$0.3, becomes a charge-ordered insulator at x=0.5,
and then to a 'Curie-Weiss metal' at x $\sim$ 0.70. The ground
state of Na$_{x}$CoO$_2$ for $x>0.75$ is still not clear. Sugiyama
and co-workers carried out the muon-spin-rotation experiments
($\mu$SR) on a Na$_{0.9}$CoO$_2$ single crystal prepared by a flux
method \cite{Sugiyama}. They found that the $\mu$SR signal was
fitted best by a zeroth-order Bessel function, and concluded that
Na$_{0.9}$CoO$_2$ undergoes a transition from a paramagnetic to
incommensurate spin density wave state (IC-SDW). They claimed that
the IC-SDW occurs within the CoO$_2$ plane, and the oscillating
moments pointed along the c-axis. However, Bayrakci et al. found
that Na$_{0.82}$CoO$_2$ exhibits a bulk antiferromagnetic (AF)
long range order with a Neel temperature about 20K by the
susceptibility, specific heat and $\mu$SR measurements for
Na$_{0.82}$CoO$_2$ single crystals prepared by floating-zone
method \cite{Bayrakci}. They found that the magnetic order
encompasses nearly 100$\%$ of the crystal volume. The inelastic
neutron scattering experiment on Na$_{0.75}$CoO$_2$ single
crystals by Boothroyd et al.  \cite{Boothroyd}showed that there
exists strong in-plane ferromagnetic (FM) correlations with an
energy scale much higher than 20K, consistent with an AF
correlation modulation perpendicular to the CoO$_2$ planes. The
existence of strong in-plane FM correlations is also consistent
with the band structure calculations \cite{Singh}.

In this Letter, we report the magnetization, specific heat and
magnetoresistance measurements on a high quality single
Na$_{0.85}$CoO$_2$ crystal in magnetic fields up to 14T. For
$H\parallel c$, we find that there is a metamagnetic transition
from an AF to a quasi-FM state at $\sim$8T below 20K. However, for
$H\perp c$, no transition is observed in magnetization
measurements up to 14T. Our results suggest that the competition
between the AF and FM correlations controls the physics of
low-lying excitations in this material.

A single crystal of Na$_{0.85}$CoO$_2$ was prepared in flowing
O$_2$ in a floating-zone optical image furnace. The starting feed
and seed materials were prepared using NaCO$_3$ and Co$_3$O$_4$
powders with Na:Co ratio of 0.85:1. The well-mixed powders were
heated at 750$^o$C overnight and then the powders were reground
and heated at 850$^o$C for a day. The mixture was then pressed to
form a cylinder of about 10cm in length and 8mm in diameter for
feed. The growth rate is 2mm per hour. The end part of this
crystal, which is the most homogeneous part of the crystal, was
cut into several pieces and used in our measurements. The Na:Co
ratio of the crystal, determined by the induction-coupled plasma
measurements, is 0.85$\pm$0.02. From the x-ray diffraction
measurements, no impure phase is detected within experimental
error of 2$\%$ and the c-axis lattice constant c is determined to
be 10.62$\AA$. The c dependence of x is consistent with the data
published by other groups \cite{Foo,Bayrakci,Sales}. Detailed
preparation procedure will be published elsewhere \cite{Wu}. The
magnetization, specific heat and resistivity measurements were
performed in Quantum Design PPMS systems. The field dependence of
thermometer and addenda was carefully calibrated before specific
heat measurement. The in-plane resistivity was measured using a
4-probe low frequency ac method.

\begin{figure}
\includegraphics[width=0.85\linewidth]{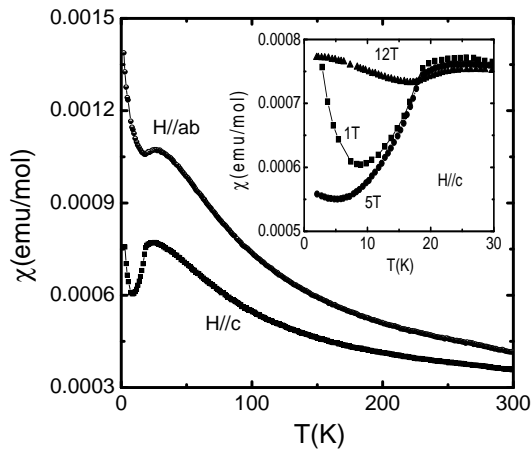}
\caption{The magnetic susceptibility $\chi = M/H$ versus
temperature $T$ of Na$_{0.85}$CoO$_2$ in a magnetic field of 1T
applied along and perpendicular to the c-axis. Inset: Low
temperature susceptibility $\chi$ versus temperature $T$ in
applied fields of 1T, 5T, and 12T along the c-axis.}
\end{figure}

Fig. 1 shows the magnetic susceptibility $\chi = M/H$ as a
function of temperature $T$ in a field of 1T applied both along
and perpendicular to the c-axis. $\chi$ exhibits a
Curie-Weiss-like behavior in high temperatures. A broad peak
appears around 27K. Below 18.5K, $\chi$ drops down sharply for
$H\parallel c$ but goes up rapidly for $H\perp c$. Similar
temperature dependence of $\chi$ was observed by Bayrakci et al.
\cite{Bayrakci} for a Na$_{0.82}$CoO$_2$ single crystal. The broad
peak around 30K can be attributed to quasi-two-dimensional AF
fluctuations. The sharp drop of $\chi$ at about 20K reveals a
phase transition from a paramagnetic to an AF state with staggered
magnetization along the c-axis. Above 50K, $\chi$ can be well
fitted by the formula $\chi(T)= \chi_0+ C/(T+\theta)$. From the
fitting parameters, the effective moments are determined to be
0.66$\mu_B$/Co for $H\parallel c$ and 0.92$\mu_B$/Co for
$H\parallel ab$. They are consistent with the data published by
Sales et al.[25] if we assume that the effective moments are
proportional to the content of Co$^{+4}$ ions in this material.

The inset of Fig. 1 shows $\chi$ as a function of temperature from
2K to 40K in different applied fields. The Curie tail in low
temperatures at 1T might be due to paramagnetic impurities. It is
suppressed at 5T. However, at 12T, $\chi$ rises up below $T_m
\sim$17.4K, and tends to saturate at low temperatures. This
indicates that a phase transition from an AF (or SDW) to a
partially FM state occurs at a field between 5T and 12T.

\begin{figure}
\includegraphics[width=0.72\linewidth]{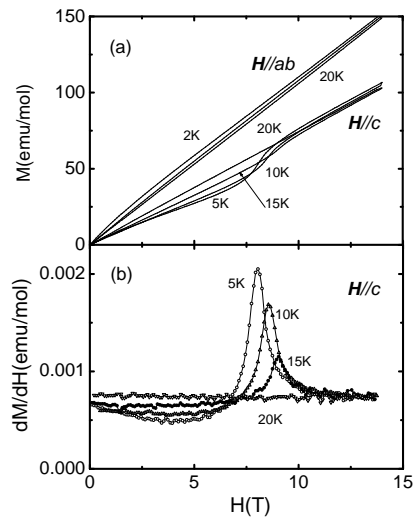}
\caption{(a) The magnetization $M$ versus field $H$ of
Na$_{0.85}$CoO$_2$ at 5K, 10K, 15K and 20K for $H\parallel c$, and
at 2K, 10K, and 20K for $H\perp c$, respectively. (b) $dM/dH$
versus $H$ at 5K, 10K, 15K and 20K for $H\parallel c$.}
\end{figure}

Fig. 2(a) shows the field dependence of magnetization $M$ at 5K,
10K, 15K and 20K for $H\parallel c$, and at 2K, 10K and 20K for
$H\perp c$. For $H\parallel c$, a rapid superlinear rise in $M(H)$
is observed at $\sim$ 8.0T for $T$= 5K. With increasing
temperatures, this superlinear rise feature is weakened and
disappears above 20K. This can be more clearly seen from the
$dM/dH$ plot [Fig. 2(b)]. The appearance of the sharp peaks in
$dM/dH$ is an indication of a phase transition. The peak position
of $dM/dH$ increases from $\sim$ 8.0T at $T$=5K to $\sim$ 9.0T at
$T$=15K. For $H\perp c$, the magnetization shows a typical
paramagnetic behavior, and $M$ increases linearly with $H$ up to
$14T$. This anisotropic magnetic response of the system suggests
that the magnetic moments of Co ions are along the c-axis,
consistent with other experiments \cite{Bayrakci, Sugiyama}.

The above results suggest that there is a metamagnetic transition around 8T in low
temperatures. Metamagnetism refers to two different kinds of magnetic phenomena. It
describes a transition or crossover from a paramagnetic state in low fields to a more
polarized state in high fields, or the spin flop transition from an AF state to a spin
ferromagnetically polarized state \cite{MackenziePRL}. The spin-flop jump here is small
compared with that in a conventional AF state. The $\mu$SR experiments indicate that
there are three kinds of local spins pointing in different directions in these materials
\cite{Bayrakci}. Probably only one kind of these local spins is involved in the
metamagnetic transition. In cuprates LaCuO$_4$ \cite{Thio} and La$_{2-x}$Sr$_x$CuO$_4$
with x=0.01 \cite{Ando}, and layered manganite La$_{1.4}$Sr$_{1.6}$Mn$_2$O$_7$
\cite{Welp}, spin flop transitions from an AF state to a ferromagnetically polarized
state were observed by magnetization measurements. The metamagnetic transition in
metallic systems were also observed in the bilayer perovskite ruthenate Sr$_3$RuO$_7$
\cite{MackenziePRL, Mackenziescience} and heavy fermion systems URu$_2$Si$_2$ \cite{Kim}
and UPt$_3$ \cite{Frings} with possible quantum critical phenomena.

\begin{figure}
\includegraphics[width=0.68\linewidth]{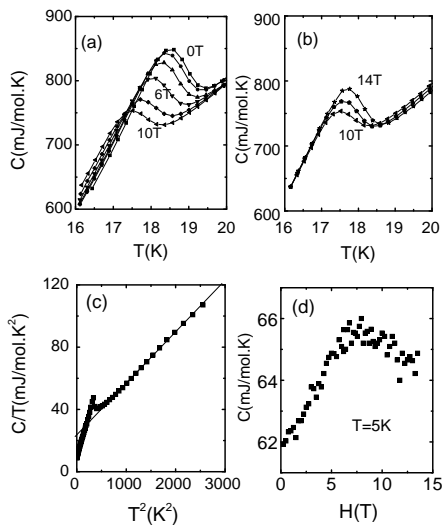}
\caption{ The specific heat $C$ of Na$_{0.85}$CoO$_2$ crystal for
$H\parallel c$ up to 14T. (a) $C$ versus $T$ in the transition
temperature regime at 0T, 2T, 4T, 6T, 8T and 10T (from the top to
the bottom), respectively. (b) Same as for (a), but at 10T, 12T
and 14T (from the bottom to the top), respectively. (c) $C/T$
versus $T^2$ in zero field. (d) $C$ versus $H$ at 5K.}
\end{figure}

The metamagnetic transition has also been observed in the field
dependent specific heat and in-plane magnetoresistance (MR)
measurements. The specific heat of Na$_{0.85}$CoO$_2$ shows a
sharp peak at about 18.5K at zero field. It corresponds to the
magnetic ordering transition as observed in the susceptibility
measurements. Figs. 3(a) and 3(b) show the specific heat $C$ from
16K to 20K in eight different fields applied along the c-axis. The
magnetic ordering transition temperature $T_m$ decreases with
increasing $H$ below 10T [Fig. 3(a)], but increases with $H$ above
10T [Fig. 3(b)]. For a magnetic transition from a paramagnetic
state to an AF state, the applied magnetic fields suppress the AF
correlations as well as the transition temperature. However, for a
transition from a paramagnetic state to a FM state, the transition
temperature increases with increasing fields since the FM
correlations are enhanced in an applied field. Our specific heat
results indicate that the magnetic ordering is AF-like in low
fields, but FM-like in high fields, consistent with the
magnetization measurement given above.

\begin{figure}
\includegraphics[width=0.74\linewidth]{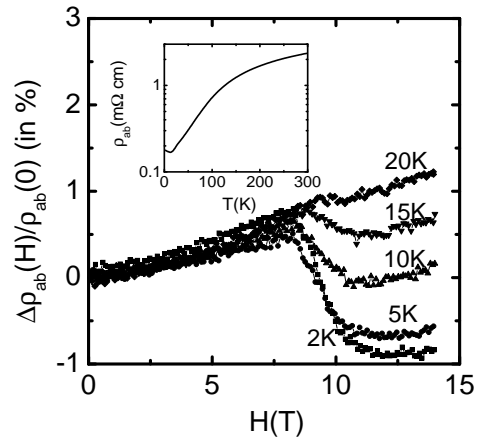}
\caption{ The in-plane magnetoresistance of Na$_{0.85}$CoO$_2$ at
five different temperatures for $H\parallel c$. Inset: The
temperature dependence of the in-plane resistivity at zero field.}
\end{figure}

Above 22K, the specific heat is well described by a sum of the
phonon and electronic contributions $C \sim \gamma T + \beta T^3$
[Fig. 3(c)]. From the fitting, $\gamma$ is found to be
$\sim$24mJ/mol.K$^2$. This value of $\gamma$ is close to the value
(27mJ/mol.K$^2$) estimated by Sales \cite{Sales} et al. and by
Motohashi et al. \cite{Motohashi} for Na$_{0.75}$CoO$_2$ crystals.
The Debye temperature $\theta _D$ deduced from $\beta$ is about
598K, slightly higher than the corresponding value 550K for
Na$_{0.75}$CoO$_2$ \cite{Sales, Motohashi}. The values of $\gamma$
and $\theta _D$ obtained by Bayrakci et al. for Na$_{0.82}$CoO$_2$
\cite{Bayrakci} differ from ours since their values were deduced
from the data below $T_m$, and the magnetic contribution was not
considered.

Fig. 3(d) shows the field dependence of $C$ at 5K for $H\parallel
c$. In low fields, $C$ increases with increasing field, but drops
down above 8T. The critical field $H_m$ obtained from the specific
heat is consistent with that obtained from magnetization
measurements. The increase of the specific heat in low fields can
be explained by the field induced enhancement of spin density of
states. The electronic specific heat is proportional to the total
density of low-lying excitations. The applied field suppresses the
AF coupling between the CoO$_2$ layers and enhances the density of
spin excitations. Above $H_m$, a transition to a ferromagnetically
polarized state occurs and the specific heat decreases with
further increasing field. When the field is large enough to
polarize all the spins, the spin excitations will be completely
suppressed and will have no contribution to the specific heat.

The inset of Fig. 4 shows the temperature dependence of the
in-plane resistivity. The in-plane resistivity exhibits a metallic
behavior above 18K. Below 18K, $\rho_{ab}$ increases with
decreasing temperature. Fig. 4 shows the in-plane MR from 2K to
20K up to 14T. In low fields, a positive MR is observed. This is
due to the suppression of the AF order and the enhancement of spin
scattering of conducting electrons by the applied field. However,
around 8T, a sharp drop in MR occurs at low temperatures. This is
because the presence of the FM order tends to suppress the spin
scattering. Similar MR behaviors across the AF and FM phase
boundary were observed in Layered Ruthenates \cite{Nakatsuji} and
CMR materials.

The above discussion shows unambiguously that the ground state of
Na$_{0.85}$CoO$_2$ is AF ordered in low fields and FM ordered in
high fields. Since the in-plane spin correlations revealed by the
neutron scattering measurements are predominately
FM\cite{Boothroyd}, this suggests that the inter-layer spins are
AF coupled but become ferromagnetic correlated after the spin flop
transition in Na$_{0.85}$CoO$_2$. Thus the metamagnetic transition
here corresponds to a spin flop transition from an AF to a FM
state along the c-axis. From the transition field, the
characteristic energy of the interlayer antiferromagntic coupling
is estimated to be $\sim 1{\rm meV}$.

\begin{figure}
\includegraphics[width=0.76\linewidth]{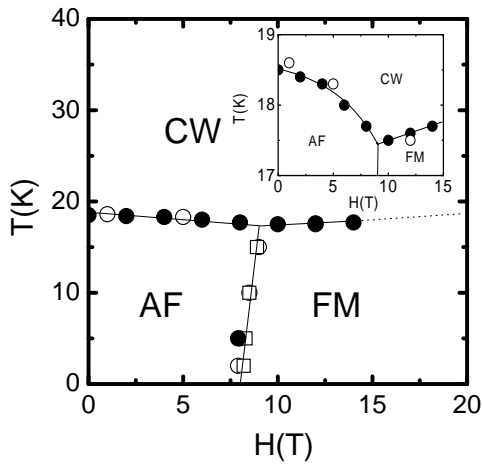}
\caption{ The magnetic phase diagram of Na$_{0.85}$CoO$_2$. The
inset shows the data around the tri-critical point in an enlarged
scale. CW, AF and FM represent a Curie-Weiss, an
antiferromagnetically ordered, a spin ferromagnetically polarized
state, respectively. The phase boundary points are obtained from
the susceptibility (open circle), specific heat (closed circle)
and magnetoresistance (open square) measurements.}
\end{figure}

From the above measurements, we can draw a low temperature
magnetic phase diagram for Na$_{0.85}$CoO$_2$. As shown in fig.5,
in higher temperatures, Na$_{0.85}$CoO$_2$ is in a
Curie-Weiss-like phase. But by lowering temperature, it falls into
either an AF phase in low fields, or into a ferromagnetically
polarized phase in high fields. The AF phase is connected with the
ferromagnetically polarized phase via a spin flop transition. For
a x=0.8$\pm$0.02 single crystal, we have observed a similar
metamagnetic transition. In this case, the zero field AF
transition temperature is 20.2K and the metamagnetic transition
occurs at about 9.5T in low temperatures. Since our measurements
data of $\chi$, $C$, and $\rho_{ab}$ behave similarly as for other
Na$_{x}$CoO$_2$ with $x>0.75$, we believe that this magnetic phase
diagram is applicable to all Na$_{x}$CoO$_2$ with x$>$0.75.

In summary, the magnetization, specific heat and magnetoresistance
were measured for a Na$_{0.85}$CoO$_2$ single crystal in applied
magnetic fields up to 14T. It is found that there is a
metamagnetic phase transition around 8T for $H\parallel c$ below
20K. No transition is observed in magnetization measurements up to
14T for $H\perp c$. The magnetic phase diagram of Na$_{x}$CoO$_2$
with x$\sim$0.85 is determined from the susceptibility, specific
heat and magnetoresistance data. In high temperatures, the system
is in a Curie-Weiss-like paramagnetic phase. However, in low
temperatures, the system can be either in an AF ordered phase in
low fields, or in a partially FM phase in high fields.

We thank L. Lu, Y. P. Wang, G. M. Zhang and Z. J. Chen for useful
discussions. This work is supported by NSFC Grants 10274101,
10025418 and 10374109.

\end{document}